\documentstyle[aps,twocolumn,epsfig]{revtex}
\begin{document}

% \preprint{
% \vbox{
% \hbox{February 2000}
% \hbox{ADP-99-47/T384}
% \hbox{JLAB-THY-99-39}
% }}

\title{Comment on ``Parton distributions, $d/u$,
	and higher twist effects at high $x$''}

\author{W. Melnitchouk$^{1,2}$, I.R. Afnan$^3$, F. Bissey$^1$
	and A.W. Thomas$^1$}

\address{$^1$	Special Research Centre for the
		Subatomic Structure of Matter,
		and Department of Physics and Mathematical Physics,
		University of Adelaide, 5005,
		Australia}
\address{$^2$	Jefferson Lab,
		12000 Jefferson Avenue,
		Newport News, VA 23606}
\address{$^3$	School of Physical Sciences,
		The Flinders University of South Australia,
		Bedford Park, S.A. 5042,
		Australia}
\maketitle

%%%%%%%%%%%%%%%%%%%%%%%%%%%%%%%%%%%%%%%%%%%%%%%%%%%%%%%%%%%%%%%%%%%%%%%%%%%%%
{\bf I.}\ \
In a recent Letter Yang and Bodek \cite{YB} presented results of a new
analysis of proton and deuteron structure functions in which the free
neutron structure function, $F_2^n$, was extracted at large $x$.
Knowledge of $F_2^n$ is crucial for determining the neutron/proton
structure function ratio, whose $x \to 1$ limit is sensitive to
mechanisms of SU(6) spin-flavor symmetry breaking, and provides
one of the fundamental tests of the $x$ dependence of parton
distributions in perturbative QCD.

Relating nuclear structure functions to those of free nucleons is, 
however, not straightforward because at large $x$ nuclear effects
become quite sizable.
In particular, omitting nuclear binding or off-shell corrections can
introduce errors of up to 50\%~\cite{MT,WHITLOW} in $F_2^n/F_2^p$
already at $x \sim 0.75$.
Rather than follow the conventional procedure of subtracting Fermi
motion and binding effects in the deuteron via standard two-body
wave functions \cite{UNSMEAR}, Yang and Bodek instead extract $F_2^n$
using ``a model proposed by Frankfurt and Strikman \cite{FS}, in which
all binding effects in the deuteron and heavy nuclear targets are
assumed to scale with the nuclear density'' \cite{YB}.
Here we point out why this approach is ill-defined for light nuclei,
and introduces a large theoretical bias into the extraction of
$F_2^n$ at large $x$.

For heavy nuclei the nuclear EMC effect is observed to scale with the
nuclear density, $\rho_A$~\cite{FS,SLAC,GST}
\begin{eqnarray}
\label{AD}
{ R_{A_1} - 1 \over R_{A_2} - 1 }
&=& { \rho_{A_1} \over \rho_{A_2} }\ ,
\end{eqnarray}
where $R_A = F_2^A/F_2^d$ and $\rho_A = 3A/(4\pi R_e^3)$,
with $R_e^2 = (5/3) \langle r^2 \rangle$ and
$\langle r^2 \rangle^{1/2}$ is the nuclear r.m.s. radius.
Assuming that an analog of Eq.(\ref{AD}) holds also for $F_2^A/F_2^N$
($F_2^N = F_2^p + F_2^n$) one finds
\begin{eqnarray}
\label{DN}
{ F_2^d \over F_2^N}
&=& 1 + { \rho_d \over (\rho_A - \rho_d\ R_A) }\ ( R_A - 1 ) .
\end{eqnarray}
This expression was derived by Frankfurt and Strikman in Ref.\cite{FS},
where the denominator was further approximated by
$\rho_A - \rho_d\ R_A \approx \rho_A - \rho_d$.
It was used in the analysis of the SLAC data \cite{SLAC}, and also
referred to by Yang and Bodek \cite{YB}, although the explicit formulas
used in that analysis are not given.
We have checked the numerical values for $F_2^d/F_2^N$ in \cite{YB},
and they agree with the result one would obtain from Eq.(\ref{DN})
if the densities quoted in \cite{SLAC} are calculated in terms of
charge radii \cite{ATOM}.

Frankfurt and Strikman point out \cite{FS} that from the above expression
for $F_2^d/F_2^N$ one can extract the free neutron structure function from
empirical EMC ratios and the nuclear densities.
With the numerical values for $\rho_A$ quoted in \cite{SLAC},
one finds then that the EMC effect in $d$ is about 25\% as large
as in $^{56}$Fe \cite{YB,FS}, and has the same $x$ dependence.

While the correlation of EMC ratios with nuclear densities is empirical
for heavy nuclei, application of Eq.(\ref{AD}) to light nuclei, $A < 4$,
for which EMC effect has not yet been determined,
is fraught with ambiguities in defining physically meaningful nuclear
densities for few-body nuclei.
Firstly, the relevant density in Eq.(\ref{AD}) is the nuclear matter 
density, while in practice $\rho_A$ is calculated from the charge
radius \cite{YB,SLAC} --- for heavy nuclei the difference is negligible,
but for light nuclei it can be significant.
Secondly, treating the deuteron as a system with radius
$\langle r^2 \rangle^{1/2} \approx 2$~fm means that one includes
{\em both} nucleons in the average density felt by one of them,
even though one nucleon obviously cannot influence its own structure.
Therefore what one should consider is the probability of one nucleon
overlapping with the other, which is simply the deuteron wave function
at the origin.
This has zero weight, however, so the only sensible definition of mean
density for the deuteron is zero.
Strictly speaking, the nuclear density extrapolation then predicts
{\em no nuclear EMC effect in the deuteron}.
%
% [ MENTION SOMEWHERE \rho IN FORMULAS SHOULD
%   BE SCALED BY (A-1)/A rho ... ??? ]

In Ref.\cite{FS} Frankfurt and Strikman argue that for heavy nuclei the
average potential energy is proportional to the average nuclear density,
and hence for $x$ below 0.5--0.6 (where nuclear Fermi motion is not
overwhelming) the nuclear EMC effect should scale with average nuclear
density.
If one applies the idea from heavy nuclei (where the assumption is known
empirically to be reasonable) to the deuteron, one finds that the EMC
effect in $d$ is $(F_2^d/F_2^N - 1) = 0.25\ (F_2^{Fe}/F_2^d - 1)$.
% For light nuclei ($A=2, 3$), where there are no data, no justification
For light nuclei ($A=2, 3$) no justification for this assumption is
provided, however, and for $x \agt 0.6$, where nuclear Fermi motion
effects become large, Frankfurt and Strikman caution that this estimate
is only a qualitative one \cite{FS}.

In a reply to our original Comment, Yang and Bodek state that
``although the notion of nuclear density for the deuteron may not be
very well defined, the value of the nuclear density for deuterium that
was used in the SLAC fit yields a similar correction for nuclear binding
in the deuteron as the estimate by Frankfurt and Strikman'' \cite{YBR}.
As explained above, not only is the notion of nuclear density for the
deuteron not very well defined, {\em it is not defined at all}.

Moreover, agreement with other calculations \cite{SLAC,MST} for the
magnitude of the deuteron EMC effect does not provide {\em a posteriori}
justification for using an ill-defined quantity like average nuclear
density for the deuteron.
One would never think of using a density extrapolation to extract the
neutron's electromagnetic form factors from quasi-elastic scattering on
the deuteron or $^3$He, for example, and there is no reason to believe
this method is any more reasonable for structure functions.

\vspace*{0.5cm}

{\bf II.}\ \	% ---------- A=3 -----------
The size of the EMC effect in the deuteron cannot be tested directly
in any inclusive deep-inelastic scattering experiment on the deuteron,
as it requires knowledge of $F_2^n$, which itself must be extracted
from deuteron data.
If, on the other hand, the EMC effect scales with nuclear density even
for the deuteron, as assumed in \cite{YB,FS}, it must also scale with
$\rho_A$ for all $A > 2$.
In particular, it must predict the size of the EMC effect in
3-body nuclei.
In fact, for $A=3$ the nuclear density extrapolation makes
quite a dramatic prediction: since the 3-body nuclear densities
calculated from the charge radii are
$\rho_{^3He} = 0.049$ fm$^{-3}$ and $\rho_{^3H} = 0.068$ fm$^{-3}$,
the EMC effect in $^3$H is 40\% larger than that in $^3$He.
This is to be compared with standard many-body calculations in terms of
Faddeev wave functions \cite{ABT} which predict $\alt 10\%$ difference
between the EMC effects in $A=3$ mirror nuclei, see Fig.1.
The $A=3$ system presents therefore an ideal case for testing the scaling
of the nuclear EMC effect to small $A$.

\begin{figure}
\begin{center}
\epsfig{figure=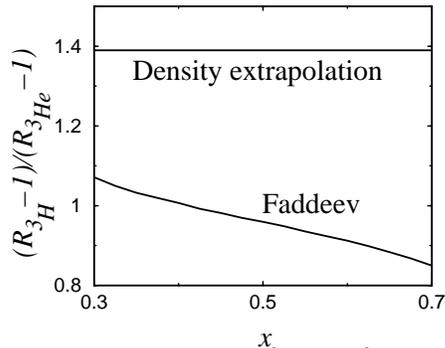,height=4.5cm}
\caption{Ratio of EMC effects in $^3$He and $^3$H using standard
	many body (Faddeev) wave functions, and the nuclear
	density extrapolation.}
\end{center}
\end{figure}

\vspace*{0.5cm}

{\bf III.}\ \	% --------- "M&T model" ---------
In Ref.\cite{YB}, Yang and Bodek compare their $F_2^d/F_2^N$ ratio with
the relativistic quark model calculation of Ref.\cite{MST}
(which we denote by ``MST'', and which is referred to in \cite{YB}
as the ``Melnitchouk--Thomas theory''),
while attributing to this the empirical extraction of $F_2^n$ which
was actually carried out in Ref.\cite{MT} (which we denote by ``MT'').
They fail to appreciate the distinction between the MST investigation
of relativistic and off-shell effects within a {\em model} and the MT
analysis of {\em data}, so a brief explanation is needed to clarify
the confusion.

In the MST model, the objective was to investigate the extent to which
relativity and nucleon off-mass-shell effects ($p^2 \not= M^2$) violate
the familiar convolution approximation for nuclear structure functions.
To test the significance of these effects, a microscopic model of nucleon
structure functions had to be constructed (in the absence of data on
off-shell nucleon structure functions!), in order to consistently
combine this with a relativistic deuteron wave function \cite{GROSS}
in a covariant calculation.
Both the proton and neutron structure functions were parameterized in
terms of a simple model for the nucleon-quark-diquark vertex functions,
and, in the context of this model, the theoretical $F_2^d/F_2^N$ ratio
was constructed.

No attempt to extract $F_2^n$ from the calculated ratio was made in the
MST model.
Indeed, if any input neutron structure function is used to construct the
theoretical $F_2^d/F_2^N$ ratio, clearly the calculated ratio cannot then
be applied to draw conclusions about the extracted $F_2^n$ --- different
neutron input would result in a different EMC ratio, and the argument
would be cyclic.
Therefore comparing the ratio calculated in the MST model with a ratio
used in an empirical extraction of $F_2^n$ such as in Ref.\cite{YB} is
rather meaningless.

The least ambiguous procedure is to deconvolute the smeared neutron
structure function by iterating the inversion procedure until a
convergent, self-consistent solution is obtained.
This was done in the MT analysis, in which the {\em only} theoretical
input was the deuteron wave function.

In Ref.\cite{YBR}, Yang and Bodek criticize the calculated $F_2^d/F_2^N$
ratio in the MST model\footnote{
Note that this is incorrectly attributed here to \cite{MT}, highlighting
the apparent confusion in \cite{YB,YBR} between the MST model and the MT
analysis.
The curve labeled ``Melnitchouk-Thomas model'' in Fig.1 of \cite{YB} is
also incorrectly referenced to \cite{MT} rather than to the MST model
\cite{MST}.}
on the grounds that its small-$x$ behavior is opposite to that in
the EMC ratio for iron.
They also assert that the small-$x$ behavior reflects some sort of
violation of the energy--momentum sum rule in the MST model.
The history of this discussion is rather long and illustrious, and for the
background we refer the reader to any standard review of the EMC effect
(see e.g. Refs.\cite{GST,BT} and references therein).
Both of these points are actually diversions and irrelevant to our main
comment, however, since they were raised as apparent justifications for
using the density extrapolation model, we feel a need to address them.

Firstly, the behavior of structure functions at $x \sim 0.2$ is clearly
outside the region of large $x$ where the nuclear effects which we are
discussing are relevant.
The focus of the MST investigation was specifically nuclear effects
in the valence quark region at large $x$.
A realistic description of the sea requires much more sophistication
than can be reasonably demanded of any simple valence quark model.
Secondly, in any realistic model of nuclear structure which properly
accounts for nuclear binding, nucleons alone obviously cannot carry
all of the momentum of the nucleus.
An analog of demanding that nucleons saturate the momentum sum rule
would be to demand that valence quarks alone carry all of the nucleon's
momentum and gluons carry none --- in contradiction with experiment.
This is the content of the assertion \cite{YBR} that the energy--momentum
sum rule is not satisfied in the MST model.

The point is that we have never advocated using the MST model to
extract the neutron structure function, contrary to the claim made
in \cite{YBR}.
As explained above, the least ambiguous $F_2^n$ extraction requires
employing the iterative procedure for unsmearing the effects of the
deuteron wave function, as outlined in \cite{MT,WHITLOW}, using deuteron
and proton {\em data} as input.
However, {\em even if} these points were valid, they would still not be
relevant to the issue of $F_2^n$ extraction as they refer specifically
to the MST model and {\em not} to the $F_2^n$ extraction in the MT
data analysis.

Yang and Bodek in addition claim \cite{YBR} that the MST model has not
been applied to $A \geq 4$ nuclei.
Since the MST model was relativistic (namely, it included terms beyond
order $v/c$ in the deuteron wave function), applying it to other nuclei
is straightforward once relativistic wave functions are known.
To date only non-relativistic wave functions have been calculated for
nuclei with $A \geq 4$.
However, since the MST model has a well-defined non-relativistic limit
\cite{PMT} (namely, omitting deuteron $P$-waves, and dropping terms of
order $v^2/c^2$), it smoothly matches onto previous non-relativistic
calculations of the nuclear EMC effect for $A \geq 4$ \cite{GST,BT}.
So again, contrary to the assertion by Yang and Bodek \cite{YBR},
the MST model has indeed been tested for all nuclei and for all cases
where wave functions exist \cite{CODE}.

\vspace*{0.5cm}

{\bf IV.}\ \	% ------------ Summary -----------
To summarize, we have demonstrated that the nuclear density model of the
EMC effect, extrapolated well beyond its region of validity to the case
of the deuteron, is a completely unreliable method of extracting the
neutron structure function at large $x$, and introduces a large
theoretical bias into the extraction procedure.
Although a nuclear density fit may be quite useful for heavy nuclei
where data exist \cite{SLAC}, its extrapolation to $A \leq 3$ where
there are no data is highly speculative, as illustrated by the
difficulty in defining physically meaningful densities for few-body
systems.

We believe this issue can only be resolved by measuring the nuclear
EMC effect in the deuteron and in $A=3$ nuclei.
One proposal \cite{H3} would be to simultaneously measure the structure
functions of $^3$He and $^3$H, extracting the $F_2^n/F_2^p$ ratio
through the cancellation of nuclear effects, which would indirectly
determine the EMC effect in the deuteron.
Other solutions would be to reconstruct $F_2^n$ from the $d/u$ ratio
extracted from parity-violating $\vec e p$ scattering \cite{PARITY},
completely free of nuclear effects, semi-inclusive $\pi^\pm$ production
from an $^1$H target \cite{PI}, or from various charged-current reactions
\cite{W}.
The above discussion presents a strong case for performing these
experiments as soon as possible.

%%%%%%%%%%%%%%%%%%%%%%%%%%%%%%%%%%%%%%%%%%%%%%%%%%%%%%%%%%%%%%%%%%%%%%%%%%
% \acknowledgements
We would like to acknowledge stimulating discussions with %J. Gomez and
G. Petratos concerning the nuclear corrections in $A=3$ nuclei.
This work was supported by the Australian Research Council.

%%%%%%%%%%%%%%%%%%%%%%%%%%%%%%%%%%%%%%%%%%%%%%%%%%%%%%%%%%%%%%%%%%%%%%%%%%

\end{document}